\documentclass{aa}  

\makeatletter
\let\linenumbers\@gobble
\let\endlinenumbers\@empty
\makeatother

\usepackage[referable]{threeparttablex}
\usepackage{todonotes}

\usepackage{longtable}
\usepackage{lscape}

\usepackage{graphicx}
\usepackage{txfonts}
\usepackage[normalem]{ulem}
\usepackage{breqn} 

\usepackage{txfonts}
\usepackage{hyperref}
\usepackage{txfonts}

\usepackage{soul}
\usepackage{orcidlink}

\begin{document}

   \title{Catalogue of central stars of extragalactic planetary nebulae}

  \subtitle{}

   \author{W. A. Weidmann\inst{1,}\inst{2}\orcidlink{0000-0002-1682-9660}
          \and
          M. B. Mari\inst{1,}\inst{2}\orcidlink{0000-0002-2363-6568}
          \and
          R. A. Pignata\inst{1,}\inst{2}\orcidlink{0000-0002-7227-9341}
          \and
          M. M. Miller Bertolami\inst{2,}\inst{3}\orcidlink{0000-0001-8031-1957}
          \and 
          E. O. Schmidt\inst{1,}\inst{2,}\inst{5}\orcidlink{0000-0001-6048-9715}
          \and
          K. Werner\inst{4}\orcidlink{0000-0002-6428-2276}
           }

   \institute{Universidad Nacional de Córdoba, Observatorio Astronómico de Córdoba, Laprida 854, X5000BGR Córdoba, Argentina \\
              \email{walter.weidmann@unc.edu.ar}
         \and
             Consejo Nacional de Investigaciones Científicas y Técnicas (CONICET), Godoy Cruz 2290, CABA, CPC 1425FQB, Argentina
%
         \and
             Instituto de Astrofísica de La Plata, Consejo Nacional de Investigaciones Científicas y Técnicas Avenida Centenario (Paseo del Bosque) S/N, B1900FWA La Plata, Argentina
         \and
             Institut für Astronomie und Astrophysik, Kepler Center for Astro and Particle Physics, Eberhard Karls Universität, Sand 1, 72076 Tübingen, Germany    
        \and
             Instituto de Astronomía Teórica y Experimental (IATE, CONICET-UNC), Laprida 854, X5000, Córdoba, Argentina
             }

   \date{}

 
  \abstract
   {Central stars of planetary nebulae (CSPNe) are essential for understanding the final evolutionary stages of 
   low- and intermediate-mass stars. 
   However, their study in extragalactic environments remains challenging due to their 
   intrinsic faintness and the limited availability of high-quality data.}
 %
   {We aim to provide a comprehensive and up-to-date catalogue of extragalactic CSPNe in order to 
   enable a more complete view of their physical properties across different galactic environments and metallicities.}
   %
   {The catalogue was assembled using data from the most recent and reliable literature sources. 
   Priority was given to collecting effective temperatures and luminosities that have either been directly reported or consistently derived. 
   When available, spectral types or specific spectral features 
   --~such as P-Cygni profiles or broad H$\alpha$ emission lines~-- were also included. 
   This approach allowed for broader characterisation of the sample, even when accurate classifications are not available.  }
 %
   {We present a new compilation of extragalactic CSPNe --~the largest to date~-- 
   comprising over 800 objects located in the Magellanic Clouds, NGC~300, NGC~5128, and fifteen other nearby galaxies. 
   This catalogue enables, for the first time, a global comparison of CSPNe physical parameters beyond the Milky Way. 
   Updated Hertzsprung–Russell diagrams are provided featuring CSPNe from seven different 
   galaxies, revealing trends and outliers that merit further investigation. 
   The catalogue represents a valuable resource for future spectroscopic 
   follow-up and for improving our understanding of post-AGB evolution in diverse galactic contexts.   }
   {}

   \keywords{Catalogs --
                planetary nebulae: general --
                stars: evolution --
                Local Group
               }

   \maketitle
%


\section{Introduction}

Planetary nebulae (PNe) represent a brief 
phase in the evolution  
of low- and intermediate-mass stars, i.e. those with masses between 0.8 and 8.0
M$_{\odot}$. 
Within the standard single stellar evolution scenario, 
PNe are formed when the progenitor star ejects most of its envelope and the remaining stellar remnant contracts and heats up. 
During this stage, the previously ejected material is ionised by the central star, 
producing the characteristic glowing structure \citep{2007oepn.book.....K,2022PASP..134b2001K}. 
The central stars of planetary nebulae (CSPNe) play a fundamental role in shaping the morphology and evolution of PNe. 
Additionally, they serve as valuable probes of stellar evolution, providing insight into the properties of stellar
populations and the chemical composition of their host environments.

In the Milky Way (MW), extensive efforts have been made to catalogue
PNe and their central stars.
A key contribution to the latter is the compilation by 
\citet{2011A&A...526A...6W,2020A&A...640A..10W,2021A&A...656A..51G}, 
which made it possible to assess the detectability of CSPNe. 
That catalogue showed that only about $\sim$21\% of CSPNe can be identified and characterised 
in terms of their properties, such as temperature and luminosity, 
due to observational challenges and intrinsic faintness. 
Additionally, \citet{2020A&A...640A..10W} found that 50\% of the sample 
(88 out of 175 objects) with derived values of luminosity ($L$), effective temperature ($T_{eff})$, 
and surface gravity ($g$) have masses and ages consistent with single stellar evolution models, 
suggesting that some single stars are indeed capable of forming PNe.

Due to their brightness, PNe can be detected in other galaxies. Notably, 
\citet{1989ApJ...339...39J} and \citet{1989ApJ...339...53C} 
demostrated that PNe can serve as distance calibrators 
through the planetary nebula luminosity function (PNLF). 
Significant efforts have been devoted to identifying 
PNe in other galaxies, leading to the discovery of several thousand extragalactic PNe
\citep[e.g. ][]{2002ASPC..273...41F,2006IAUS..234....9M}.
Moreover, intracluster PNe have even been detected
\citep[e.g.][]{2007PASJ...59..419A,2018MNRAS.477.1880S}.

The study of PNe in extragalactic environments offers a complementary perspective to that of the MW, 
helping to clarify how factors such as metallicity and star formation history influence stellar evolution 
\citep{1999ApJ...515..169J,2010ApJ...714.1096S,2025arXiv250117926V}. 
Extragalactic PNe have been identified in galaxies such as the 
Magellanic Clouds \citep{2010MNRAS.405.1349R}, 
M~31 \citep{2006MNRAS.369..120M}, and M~33 \citep{2004ApJ...614..167C}, 
providing unique data on stellar evolution across different galactic environments. 
However, characterising the extragalactic CSPNe is more challenging due to their 
low luminosity and the distances involved \citep{2018NatAs...2..580G}.

Despite these challenges, the growing amount of published data opens the possibility of
consolidating existing information into a unified catalogue. 
Previous efforts, such as the most recent catalogue of CSPNe by 
\citet{2020A&A...640A..10W}, have demonstrated 
the value of systematic data collection in enabling robust statistical and comparative studies.

In this work, we present a compilation of inferred properties of extragalactic CSPNe. In particular, 
we focus on their $T_{eff}$, $L$, 
and (when available) their spectral type (SpT).
While the spectral type provides essential information about the evolutionary status of these stars, 
obtaining it is particularly challenging due to the low brightness of the targets. 
As a result, the properties of a significant part of these central 
stars --~such as $L$,  $T_{eff}$, and even spectral classification~-- 
have been inferred through modelling efforts rather than direct observations. 
This catalogue integrates data from direct measurements and models, 
offering a valuable resource for investigating how the properties of central stars 
vary across different galactic environments. 
We expect this catalogue to serve as a foundation for future observational and theoretical
studies on stellar evolution across diverse environments.

While the primary objective of this work is to provide a comprehensive compilation of candidate CSPNe, 
we acknowledge that not all objects included can be considered confirmed PNe. 
Similarly, the derived or compiled stellar parameters (e.g. effective temperature and luminosity) should be treated with caution. 
Given the limitations and heterogeneity of the available data, 
we emphasise that this catalogue is intended as a resource for exploratory analysis and 
future observational follow-up, rather than as a definitive dataset for deriving robust correlations with metallicity or progenitor mass.

The structure of this paper is as follows. In Section~\ref{colum}
we provide a detailed description of the catalogue and discuss its limitations, while
in Section~\ref{para} we analyse and discuss the distribution of the physical parameters of 
the extragalactic PNe. Finally, the discussion 
and conclusions are presented in Section~\ref{discusi} and \ref{conclu}.

   
\section{The catalogue of extragalactic CSPNe}
\label{colum}

The main objective of this work is to compile known properties of extragalactic CSPNe, 
including the spectral type (SpT), $\log(L/L_\odot)$, and $\log (T_{eff})$.
The catalogue  includes 1073 objects across 19 galaxies.
Table~\ref{gal-table} presents the main characteristics of these galaxies. 
Nearly all the galaxies included, except M~81, NGC~5128, NGC~300, and NGC~5236, are members of the Local~Group. 
This underscores the substantial volume of extragalactic PNe with available spectroscopic data.

\begin{table*}[ht]
\caption{Main parameters of the galaxies in our sample. Namely, these are galaxy name, 
number of catalogued PNe, metallicity,  morphological type, 
color, and global stellar mass.
} 
\label{gal-table}      
\centering                          
\begin{tabular}{cccccc}        
\hline
 Galaxy & \# PNe & [Fe/H]$^{(1)}$  & Morph type   & (B-V)$_T^0$     & $\log(M_g$/M$_\odot$)   \\
        &        &  dex            &(NED)$^{(2)}$ & (RC3)$^{(3)}$   &   \\
\hline                        
MW          & 206 &           -                        & SBc          &   -   & 10.78$^{(6)}$  \\
LMC         & 121 &  $-$0.31 $\pm$ 0.04$^{(7)}$        & SB(s)m       & 0.43  & 9.15$^{(4)}$   \\
SMC         & 47  &  $-$0.97 $\pm$ 0.05$^{(8)}$        & SB(s)m pec   & 0.36  & 8.59$^{(4)}$   \\
M~31        & 40  &  -                                 & SA(s)b       & 0.76  & 10.95$^{(4)}$  \\
M~32        & 10  & $-$1.1 $\pm$ 0.2                   & compact E2   & 0.88  & 9.07$^{(4)}$  \\
M~33        & 11  &  -                                 & SA(s)cd      & 0.46  & 9.61$^{(4)}$   \\
M~81        & 11  & -                                  & Sb(r)I-II    & -     & 11.36$^{(10)}$      \\
IC~1613     & 1   & $-$1.3 $\pm$ 0.2                   & IB(s)m       & 0.65  & 8.00$^{(5)}$   \\
NGC~147     & 6   & $-$1.1 $\pm$ 0.2                   & E5 pec       & 0.78  & 8.30$^{(4)}$  \\
NGC~185     & 1   & $-$1.2 $\pm$ 0.15                  & dE3p         &  -    & 7.83$^{(5)}$  \\
NGC~205     & 4   & $-$0.8 $\pm$ 0.1                   & Sph          &  -    & 8.47$^{(5)}$   \\
NGC~300     & 41  &     $-0.5^{(9)}$                   & SA(s)d       & 0.56  & 9.41$^{(4)}$   \\
NGC~3109    & 1   & $-$1.5 $\pm$ 0.3                   & SB(s)m${^c}$ & -     & 8.18$^{(4)}$   \\
NGC~5128    & 557 &    -                               & S0+S pec     &  -    &   -     \\
NGC~5236    & 2   &    -                               & Scd?         & 0.61  &   -     \\
Fornax~dSph & 1   & $-$1.3 $\pm$ 0.2                   & dE4          & -     & 7.30$^{(5)}$    \\
Phoenix     & 1   & $-$1.9 $\pm$ 0.1                   & lAm          & -     &  5.88$^{(5)}$   \\
Sextans~A   & 1   & $-$1.9 $\pm$ 0.3                   & lBm          & 0.35  &   7.64$^{(5)}$  \\
Sextans~B   & 5   & $\sim$ $-$1.2                      & lB(s)m       & 0.47  &  7.71$^{(5)}$   \\
Sgr~dSph    & 3   & $-$1.0 $\pm$ 0.2                   & dSph(t)      & -     &  7.32$^{(5)}$   \\
\hline                          
\end{tabular}

\tablebib{
 (1) \cite{1998ARA&A..36..435M}; (2) NASA/IPAC Extragalactic Database; (3) \citet{Vaucouleurs_1991rc3..book.....D}; (4) \citet{2020MNRAS.498.5367D}; (5) \citet{McConnachie_2012AJ....144....4M}; (6) \citet{2015ApJ...806...96L}; (7) \citet{2002A&A...396...53R}; (8) \citet{2020MNRAS.497.3746C}; (9) \citet{1985ApJ...298..240R};
(10) \citet{1981MNRAS.195..327A}.
  }

\end{table*}

While the catalogue provides a comprehensive list of CSPNe, 
only a few of the entries include references to spectral type, with the 
majority of these objects located in the Large Magellanic Cloud (LMC).

Data were gathered from refereed publications post 1985. 
For objects with multiple determinations, we report the most recent values. 
Information in the catalogue, organised by galaxy and right ascension (RA), includes the following:

   \begin{itemize}
      \item Col.~1: Name of the PN, prioritising the 
      designation used in the original publication from which the data was obtained.
      \item Col.~2: Host galaxy. 
      \item Col.~3-4: The RA and DEC as listed in the original publications, 
      with minimal differences from the Centre de Données astronomiques de Strasbourg (CDS).
      \item Col.~5-6: Logarithm of the CSPN's effective temperature ($\log~T_{eff}$) and its reference.
In some galaxies, data for the central star were unavailable. 
However, the emission line intensities of the PNe that could be identified have been reported in the literature. 
In these cases, we were able to estimate the temperature (see Sec.~\ref{ec-tem}).
      \item Col.~7-8: Logarithm of the luminosity, in units of solar luminosity (log~$L/L_{\odot}$), and reference source.
      In the same way as for the temperature, we have been able to compute the luminosity using the magnitude of the nebula at [O{\sc iii}]$\lambda$5007       (see Sec.~\ref{ec-tem}).
      \item Col.~9-10: Spectral type and reference. 
      In cases where the spectral type was not explicitly listed, additional information 
      (e.g. P-Cygni profiles, broad H$\alpha$ emission) helped us infer the characteristics.
      \item Col.~11: References identifying the CSPN as a binary system.
      \item Col.~12: Comments.
   \end{itemize}

The coordinates have been rounded to the nearest tenth or to `.00' where precision is lacking.
We followed the same criterion for $\log(L/L_\odot)$ and $\log (T_{eff})$.
%
While errors for the physical parameters are sometimes given in the original publications, 
they are not included in this catalogue. We provide a discussion on uncertainties in Sec.~\ref{advertencia}.


\subsection{A few caveats}\label{sec_2.1}
\label{advertencia}

As mentioned earlier, only a few extragalactic CSPNe have a spectral classification.
Detecting an O(H) star is challenging due to their faint absorption lines, 
which require a high signal-to-noise (S/N) spectra. 
In contrast, Wolf-Rayet ([WR]) stars
 are easier to identify even with a lower S/N in the continuum, 
as their strong emission lines are prominent. 
Indeed, in this catalogue, the vast majority of CSPNe with a determined spectral type are [WR].

For all reported objects, the parameters of the central star were estimated 
from the observed nebular spectra. 
While most properties were primarily inferred using {\sc Cloudy} models 
\citep[][and references therein]{1998PASP..110..761F}, in some cases, 
the Zanstra temperature was also determined \citep[e.g.][]{2004ApJ...614..716V}.
This limitation directly affects the accuracy of stellar parameter determinations. 
Moreover, in most cases, only low-quality spectra are available, thus only allowing for 
an estimation of temperature and luminosity (see Sec.~\ref{ec-tem}).
The most reliable approach to accurately obtaining physical parameters for CSPNe is 
through fitting stellar atmosphere models to their spectra
\citep[e.g.][]{1988A&A...190..113M}. 
To illustrate the uncertainties associated with the stellar parameters listed in our catalogue, 
Table~\ref{dif-para} presents examples of extragalactic CSPNe with varying determinations of these physical parameters.
In addition, in Section~\ref{error} 
we present a discussion on the uncertainties associated with two specific methods.

\begin{table*}[ht]
\caption{Some objects with different determinations in their physical parameters.}
\label{dif-para}      
\centering                          
\begin{tabular}{c c c c c c c}        
\hline\hline                 
 PN        & $\log T_1$(ref.) & $\log T_2$ (ref.) & $\log T_3$ (ref.) & $\log L_1$ (ref.) & $\log L_2$ (ref.) & $\log L_3$ (ref.)  \\
\hline                        
LMC-SMP~35 & 5.07 (DM91c) & 5.03 (DV97) & 4.78 (HB04) & 3.17 (DM91c) & 3.38 (DV97) & 3.01 (HB04) \\
LMC-SMP~67 & 4.64 (DM91c) & 4.74 (HB04) & 4.69 (VS07) & 3.33 (DM91c) & 3.70 (HB04) & 3.40 (VS07) \\
LMC-SMP~85 & 4.62 (DM91c) & 4.63 (DM91) & 4.60 (HB04) & 3.85 (DM91c) & 3.79 (DM91) & 3.41 (HB04) \\
\hline                          
\end{tabular}

\tablebib{
DM91:   \cite{1991ApJ...367..115D};
DM91c:  \cite{1991ApJ...377..480D};
DV97:   \cite{1997ApJ...474..188D};
HB04:   \cite{2004ApJ...611..294H};
VS07:   \cite{2007ApJ...656..831V}.
  }
\end{table*}

Another crucial aspect to consider is that the objects in this catalogue
should be regarded as PN candidates, even if the paper does not explicitly state this. 
These objects are poorly studied, and in most cases there is only a single spectrum available.
Especially considering that PNe are perhaps one of the most difficult objects to classify, their status only as candidates should not be overlooked 
 \citep{2010PASA...27..129F}.

Additional considerations must also be addressed.
\cite{1994A&A...282..586M} proposed an empirical relationship between 
the temperature of the ionising source ($T_*$) in a photoionised gaseous 
nebula and the highest observed ionisation potential ($\chi_i$):
\begin{equation}\label{txi}
          T_*   = \chi_i \times 1000 \hspace{1cm} ¨[K/eV].  
\end{equation}

\noindent In general, extragalactic PNe are identified by means of the interferential filter centred on 
[O{\sc iii}]$\lambda$5007.
In this context, the temperatures of the extragalactic CSPNe of our catalogue  will be higher than 35100~K.

Another point to consider is that extragalactic PNe are detected as point sources 
due to the great distance separating us from them, which prevents their actual extent from being determined and 
makes it impossible for their surface brightness to be discerned. 
In this sense, what is actually detected is their integrated brightness. 
A reasonable assumption that only the brightest PNe are detected. 
This, in turn, implies that these PNe have bright central stars. 
Additionally, their nebulae are not too diffuse, making them optically 
thick and thus favouring their detection.
Moreover, according to  \citet{2005ApJ...629..499C}, 
the progenitors of the PNe that populate the top
0.5 mag of the PNLF (i.e. the brightest PN)
have core masses of $\gtrsim$ 0.6 M$_{\odot}$. 
Nevertheless, we cannot infer realistic masses with the 
data currently available and collected in this catalogue.
In summary, in the extragalactic context, only the most 
luminous CSPNe in compact and optically thick nebulae will be detectable.

Finally, given that all the compiled candidate PNe must be relatively bright, 
it is expected that they are in an early stage of CSPNe evolution. 
Specifically 
they are in the flat region of their evolutionary tracks, 
which is before they begin 
to lose luminosity as they cool down 
(see the evolutionary tracks of \citet{2016A&A...588A..25M}).


\section{Analysis of physical parameters}
\label{para}

\begin{figure}[ht]
\centering
\includegraphics[width=0.92\linewidth]{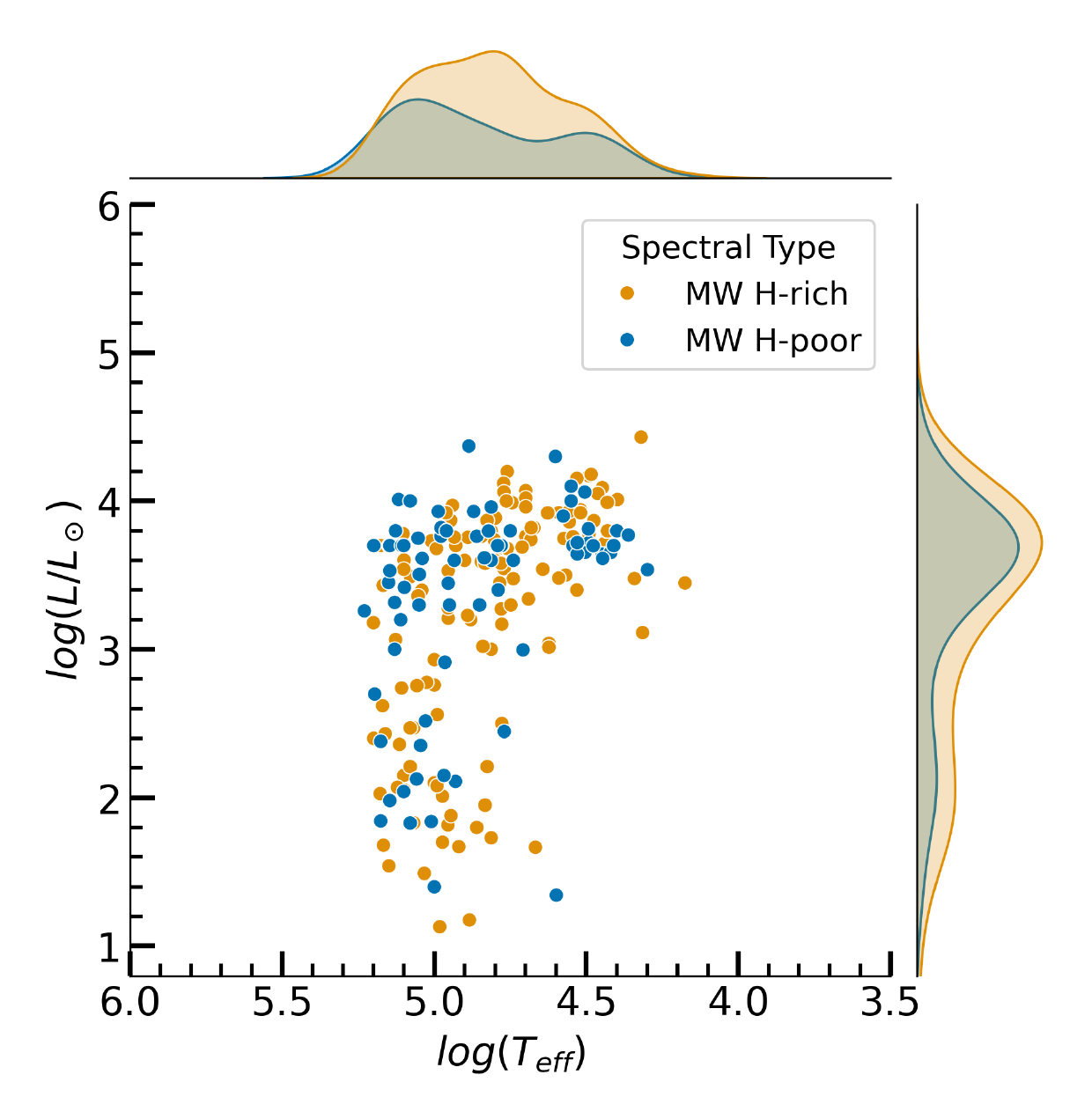}
\caption{Hertzsprung–Russell diagram showing CSPNe within the MW from \citet{2020A&A...640A..10W}. 
The marginal curves represent the data using a continuous probability density curve.}
\label{mk}
\end{figure}

To compare the distribution of the physical parameters of our CSPN sample 
with those of the MW, we used the data 
from the \cite{2020A&A...640A..10W} catalogue. 
The objects that are known to be binary were not considered.
Fig.~\ref{mk} shows the HR diagram, and H-rich and H-poor CSPNe appear to be evenly intermingled.
Consequently, in the following figures, no distinction is made between H-rich and H-poor.
It is worth noting that the physical parameters of CSPNe in our 
Galaxy were generally determined using stellar atmosphere models, 
whereas for extragalactic CSPNe, this method is unfeasible
(see Sec.~\ref{advertencia}).

\subsection{LMC and SMC}\label{lmc}

\begin{figure}[ht]
\centering
\includegraphics[width=0.92\linewidth]{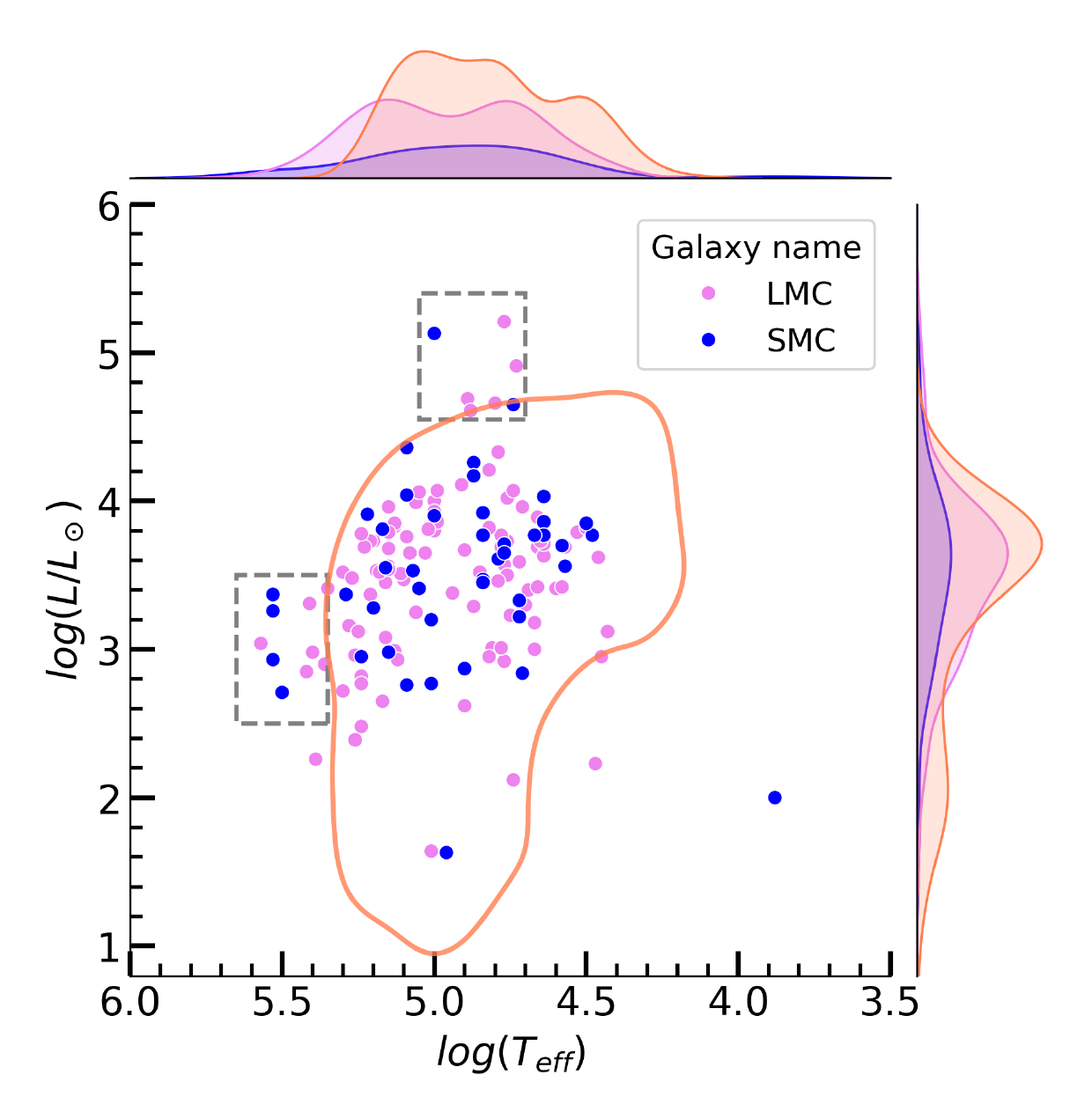}
\caption{Hertzsprung–Russell diagram of CSPNe in the LMC and SMC. 
The coldest CSPN in the SMC likely has poorly constrained parameters, 
as an effective temperature (T$_{\rm eff}$) above 10,000 K is required to efficiently ionise hydrogen.
The orange contour shows the distribution of points for the MW. In grey, 
two regions with extreme values are highlighted (see text for details).} 
\label{mw--lmc}
\end{figure}

Figure~\ref{mw--lmc} shows $\log(T_{eff}$) versus $\log(L/L_{\odot})$
for the LMC and Small Magellanic Cloud (SMC) galaxies 
along with the sample of CSPNe from the MW previously described. 
This is the most complete HR diagram of CSPNe to date for these two galaxies.
The data for the LMC and SMC
 were collected from eleven different articles, resulting in variations 
 in the methods used to derive the physical parameters. 
 We note that this may affect the comparison between the different samples 
 in Fig.~\ref{mw--lmc} given that the physical parameters were not determined using the same methodologies.

Figure~\ref{mw--lmc} highlights that the point distributions of the 
LMC and SMC are well mixed. 
Although the metallicities of the two galaxies (Table~\ref{gal-table}) 
are quite different, the quality of the data does not allow us to draw any 
conclusion about a possible trend with metallicity.

Nevertheless, the CSPNe of our most important satellite galaxies 
occupy the same region of the HR diagram as the CSPNe in our own Galaxy.
Although the point cloud of the MW is somewhat shifted towards lower temperatures 
compared to the points in the LMC and SMC, this is likely attributable to 
the methodology used for the detection of extragalactic PNe 
(see comments related to Eq.~\ref{txi}).
Additionally, we note that we did not detect PNe with 
white dwarfs (WD) nuclei.
This makes perfect sense, as PNe with WD nuclei are objects with 
low intrinsic luminosities.
Notably, the most luminous  CSPNe candidates are found in the Magellanic Clouds (Fig.~\ref{mw--lmc}).

In Fig.~\ref{mw--lmc} two regions in the HR diagram that feature 
high values of temperature and luminosity are indicated. 
While some objects in these regions are classified as probable PNe or confirmed binaries, others are not. 
In the high-luminosity region, these sources are unlikely to be CSPNe. 
According to \citet{2016A&A...588A..25M}, the most luminous CSPNe reach up to 
$\log(L/L_{\odot}) \simeq 4.4$, but even in such cases, they would 
correspond to massive CSPNe with extremely rapid evolution, making them very difficult to detect. 
Alternatively, these objects might instead be classical 
WR stars \citep[see Fig.~2 of][]{2024arXiv241004436S}.

As for the objects located in the high-temperature region, they could potentially be CSPNe. Incidentally,
they would be among the most massive ones.
Such a population of very massive CSPNe could be the progeny of high mass stars 
formed recently ($<1$ Gyr) in the Magellanic Clouds \citep{2009AJ....138.1243H, 2021MNRAS.508..245M}.
Again, such objects evolve quickly, making them difficult to detect. 
In any case, the sources in this region should be studied in more detail, 
as we might be looking at a very peculiar CSPN.


\subsection{The case of NGC~300}\label{300}

\citet{Soemitro_2023A&A...671A.142S}
obtained spectroscopic data from a moderate sample of PN in NGC~300.
This allowed them to obtain an estimate for the temperature and luminosity of the CSPNe.

The effective temperatures of the CSPNe in NGC 300 were estimated 
using the empirical excitation class (EC) method defined by \citet{2010PASA...27..187R} 
as follows:

\begin{equation}\label{eq1}
   EC_{low}=0.45  \left[ \frac{F(5007)}{F(H\beta)} \right]
\end{equation}

\begin{equation}\label{eq2}
   EC_{high}=5.54 \left[ \frac{F(4686)}{F(H\beta)} + \log_{10} \frac{F(4959)+F(5007)}{F(H\beta)}  \right],
\end{equation}
\noindent where $EC_{low}$ applies for $0<EC<5$ and $EC_{high}$ for $5\leq~EC<12$. 
Therefore, the empirical relationship between the excitation class and effective temperature is 
\begin{equation}\label{eq3}
\log (T_{eff})= 4.439 + [0.1174 (EC)] - [0.00172 (EC^2)].
\end{equation}

\noindent A peculiarity of this temperature estimation method is the discontinuity between Eqs.~\ref{eq1} and \ref{eq2}. 
As a result, applying it to a large sample of objects may produce an artificial gap in the temperature distribution.
It is important to mention that these equations are valid if the nebula is optically thick.
In addition, the datasets used did not cover the wavelength range including the 
He~{\sc ii}~$\lambda$4686 line, which is essential for determining medium- and high-excitation classes. 
As a result, the effective temperatures (T$_{\sc eff}$) for 
excitation classes greater than 5 ($EC\geq5$) could only be estimated as lower limits, with $\log (T_{\sc eff})\gtrsim5$. 
Conversely, for objects with $EC<5$, $\log (T_{\sc eff})<5$ is expected.

On the other hand, for luminosity estimates, \citet{Soemitro_2023A&A...671A.142S} assumed that 
$\sim$11\% of the CSPN's luminosity is efficiently converted into the
[O{\sc iii}]$\lambda$5007 emission line. 
This implies that a significant fraction of the CSPN's energy is channeled 
into this specific line \citep[e.g. ][]{1992ApJ...389...27D,2010A&A...523A..86S}. 
This assumption is as follows:

\begin{equation} \label{LF}
  L_{5007} = 4 \pi d^2 F_{5007},
\end{equation}

\begin{equation} \label{0.11}
  L_{5007} = 0.11 L_*.
\end{equation}

\noindent This assumption holds for optically thick nebulae and when 
[O{\sc iii}]$\lambda$5007 is the only coolant. 
If the nebula is optically thin, the efficiency of this ion decreases, 
meaning the estimated luminosities represent lower limits, similar to 
T$_{eff}$, thus making it difficult to impose stricter constraints on CSPN masses.
In this sense, \citet{Soemitro_2023A&A...671A.142S} combines Eq.~\ref{LF} and \ref{0.11}.

The proportionality constant of Eq.~\ref{0.11} depends on factors such as 
the CSPN mass and the (O/H) abundance \citep{2010A&A...523A..86S}
as well as whether it is hydrogen-burning or helium-burning \citep{1989ApJ...339...39J}
and the T$_{\sc eff}$ \citep{1992ApJ...389...27D}.

In summary, as outlined in \citet{Soemitro_2023A&A...671A.142S}, 
this represents a lower limit for luminosity and effective temperature. 
This premise is reflected in the HR diagram of Fig.~\ref{mw-5128}, where many CSPNe fall 
in the low-L ($log (L/L_{\odot})<2.5$) and low-T$_{eff}$ ($log(T_{eff})<5.1$) region, 
where the nebulae have nearly disappeared and only white dwarfs are observed 
\citep[see Fig.~9 of][]{2020A&A...640A..10W}.


\subsection{The case of M~81, NGC~185, NGC~3109, 
and NGC~5128}\label{ec-tem}

The emission line intensities of numerous PNe in some well-studied galaxies have been reported in the literature: 
M~81 \citep{2010A&A...521A...3S}, NGC~185 \citep{2008ApJ...684.1190R},
NGC~3109 \citep{2017A&A...601A.147F}, 
and NGC~5128 \citep{2015A&A...574A.109W}.
Using the [O~{\sc iii}] and He~{\sc ii} lines, we calculated the EC to estimate 
the effective temperature of the CSPNe, following the method applied by 
\citet{Soemitro_2023A&A...671A.142S}, 
Eq.~\ref{eq1}, \ref{eq2}, and \ref{eq3}.
This method requires the nebula to be optically thick.
To determine if the nebulae is optically thick, 
we used the criteria as proposed by \cite{Soemitro_2023A&A...671A.142S}: 

\begin{equation}\label{eq4}
F(6584) / F(H\alpha) > 0.3.
\end{equation}

For M~81 and NGC~185, we excluded optically thin objects from our catalogue. 
Unfortunately, the intensity of the [N~{\sc ii}]$\lambda$6584 line for objects in NGC~5128 has not been published.
In this case, as we do not know the optical depth of the PNe in the sample, we 
assumed that they are all optically thick. 
Similarly, although the line intensities of PNe in NGC~3109 have been published 
\citep{2017A&A...601A.147F}, these objects were excluded from our catalogue due to their optically thin nature.

As long as the 
[O~{\sc iii}]$\lambda$5007
and H$\beta$ lines are visible, the EC can be calculated 
through both Eq.~\ref{eq1} and \ref{eq2}. 
To decide which to apply, we adopted the following criterion: 
If emission from He~{\sc ii}$\lambda$4686 is reported, 
we use Eq.~\ref{eq2}; otherwise, we apply Eq.~\ref{eq1}.
To compute the EC of the PNe in NGC~5128, we used the approximation F(5007) = 3.01 F(4959) 
\citep{1989A&AS...80..201A}, as the intensity of the 4959\AA\ line is not reported for these PNe.

On the other hand, according to the Eq.~\ref{txi}, a temperature of 54400~K is required to 
produce He~{\sc ii} emission. 
Moreover, Eq.~\ref{eq3} indicates that an EC greater than $\sim$2.63 
is necessary for He~{\sc ii} emission.
Thus, if Eq.~\ref{eq1} yields an EC lower than 2.6 but He~{\sc ii} 
emission is observed, we attribute this emission to the central star and mark it 
as "[WR]?".

Three objects in NGC~5128 (namely, P04\_4609, EMMI\_480, and EMMI\_2045) exhibit an EC > 12, 
which we interpret as indicative of stellar emission. 
Although these objects are undoubtedly very hot, we did not assign them a specific temperature. 
Instead, we classified these PNe as [WR] candidates.

Meanwhile, for the identification of extragalactic PNe, 
images are often captured using interferential filters, 
particularly those centred on the prominent nebular line of 
[O~{\sc iii}] at 5007\AA. 
This approach is useful for estimating the luminosity of the central stars. 
It allows for the determination of PN magnitudes in this band ($m_{5007}$), 
which can then be used to estimate the central star's luminosity.
The integrated monochromatic flux from the [O~{\sc iii}]$\lambda$5007 ($F_{5007}$) 
is related to the $m_{5007}$ magnitude according to the expression given by \citet{1989ApJ...339...39J}

\begin{equation} \label{mF5007}
    m_{5007} = -13.74 - 2.5 \ log_{10} (F_{5007}), 
\end{equation}
\noindent where the flux is expressed in units of $erg \ cm^{-2} s^{-1} \AA^{-1}$.

Equation~\ref{mF5007} combined with Eq.~\ref{LF}
and Eq.~\ref{0.11}
 gives the CSPN luminosity.
 We used this set of equations to determine the luminosity of CSPN  for the 
 PNe in galaxies  M~81, NGC~185, 
and NGC~5128.

As mentioned in Sec~\ref{300}, this method of determining luminosity
applies if the nebula is optically thick
 and that [O~{\sc iii}]$\lambda$5007 is the only cooling mechanism.
For NGC~5128, we assumed that all nebulae in the sample are optically thick.

The magnitudes, $m_{5007}$, of the PNe in NGC~5128 have been 
 compiled from the ESO Multi Mode Instrument catalogue \citep[EMMI, ][]{2015A&A...574A.109W} and \cite{1993ApJS...88..423H}.
For objects with magnitudes available in both works, we applied flux averaging.

We assumed distances of 3.89~Mpc for NGC~5128 
\citep{2024MNRAS.527.5767B},
 3.63~Mpc for M~81 \citep{2001ApJ...553...47F},
and  0.616~Mpc for NGC~185  \citep{2010ApJ...711..361G}.
Additionally, we adopted 
a solar luminosity of L$_{\odot}= 3.826 \times 10^{33} erg \ s^{-1}$
\citep{1996imsa.book.....O}. 
The resulting $\log(L/L_{\odot}$), together with log($T_{eff}$) values are presented in our catalogue.

\begin{figure}[ht]
\centering
\includegraphics[width=0.92\linewidth]{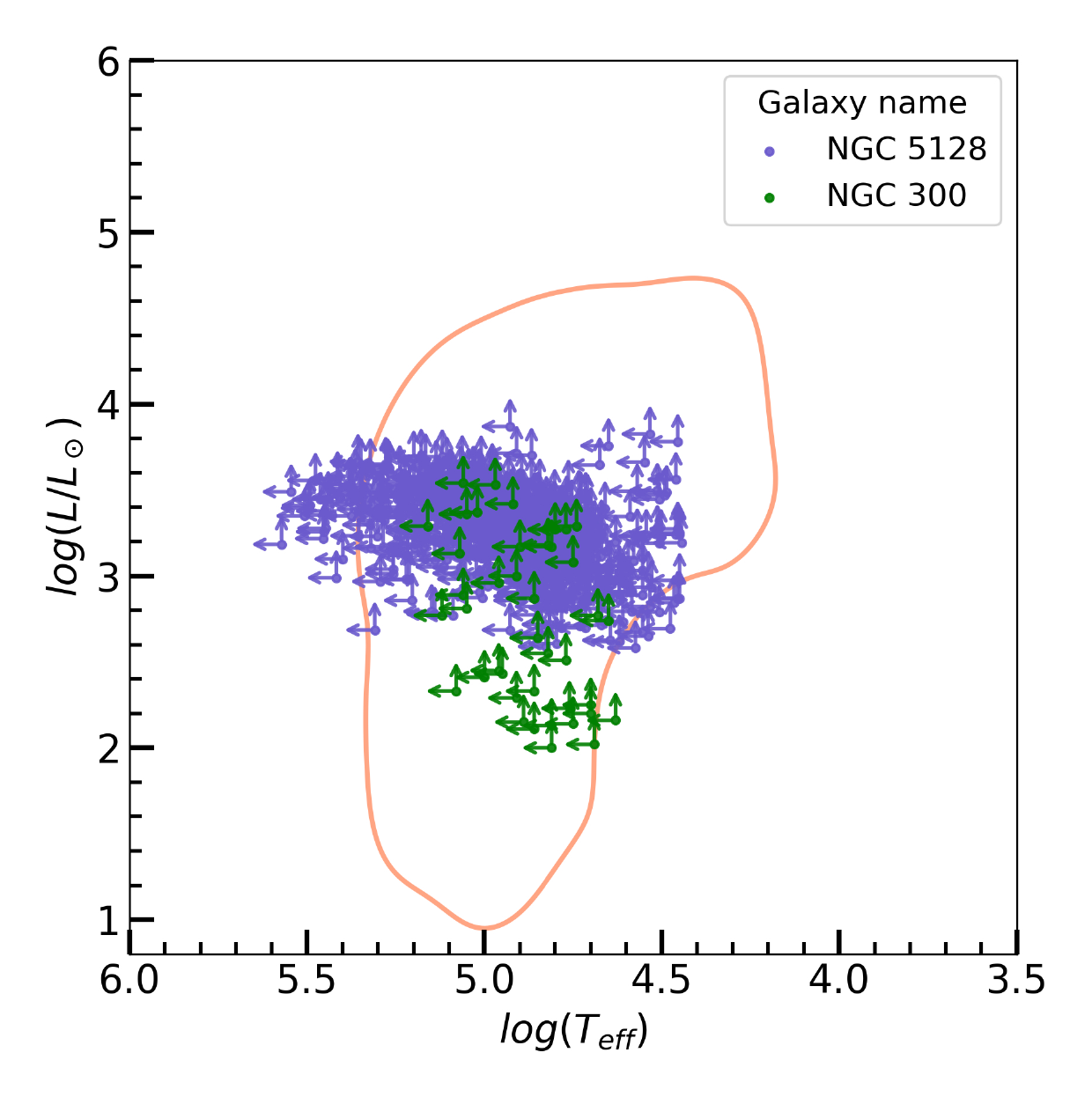}
\caption{Hertzsprung–Russell diagram for NGC~5128 and NGC~300. 
The values were calculated using the equations defined in Sec.~\ref{300}.} 
\label{mw-5128}
\end{figure}

The HR diagram is shown in Fig.~\ref{mw-5128} along with the values for NGC~300 reported by 
\citet{Soemitro_2023A&A...671A.142S}, as both datasets were derived using the same method. 
Most of the NGC~5128 points fall within the region defined by the MW 
despite the approximations made to infer their parameters. 
The distribution does not include objects in the high luminosity CSPNe region, 
most likely due to the underestimation of the stellar luminosity as a consequence of the optically thick assumption.
On the other hand, when considering their effective temperatures, we find 
--~similarly to what is observed in the Magellanic Clouds~-- 
that some objects reach high temperatures (see Sec.~\ref{lmc}).


\subsection{Local Group spiral galaxies}\label{andromeda}

\begin{figure}
\centering
\includegraphics[width=0.92\linewidth]{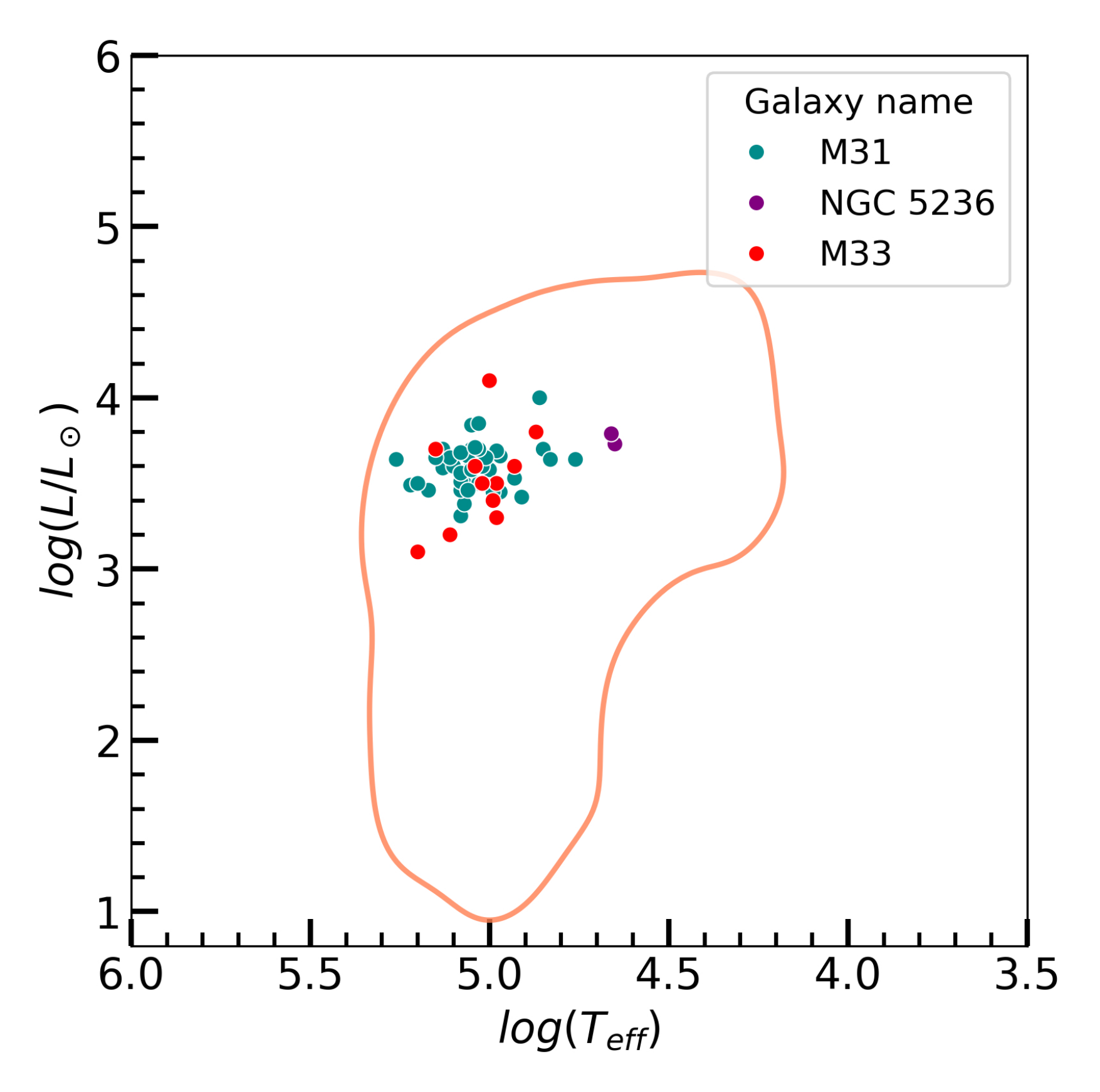}
\caption{Hertzsprung–Russell diagram for the spiral galaxies. } 
\label{mw-spy}
\end{figure}

Figure~\ref{mw-spy} shows the HR diagram of the local group spiral galaxies, in particular for M~31 and M~33. 
Given the limited sample size, our ability to draw meaningful conclusions is restricted. 
In any case, the points lie within the region defined by the CSPNe of the MW; that is, they do not exhibit anomalous values as seen in the cases of the LMC, SMC,  NGC 300, and NGC 5128. 
Nevertheless, the distribution of points appears to be tightly clustered in a region characterised 
by luminosities of log(L/$L_{\odot}) \sim$ 3.2–3.8 and effective temperatures of log(T$_{\rm eff}) \sim$ 4.8–5.2.
A possible explanation for this concentration could be a bias introduced by the 
method used to infer the stellar parameters or by the spatial distribution of the 
PNe within their host galaxies. However, the data come from four different authors, making a methodological bias unlikely. 
Furthermore, an inspection of the PNe positions shows that they are not confined 
to a specific region but rather randomly distributed throughout the galaxies.

\subsection{Assessing the accuracy of CSPN parameter estimates} \label{error}

The expressions used to estimate luminosity and temperature 
(Eqs. \ref{eq3} and \ref{0.11}) are affected by several assumptions. 
In this sense, these determinations are only meaningful in a statistical context 
involving a significant number of objects, as is the case for NGC~300 and NGC~5128.

The expression used to derive the temperature from the excitation class (Eq. \ref{eq3}) is an empirical relation.
In \citet{2010PASA...27..187R}, the authors analyse (considering the uncertainties involved)
this temperature determination method and compare it with the Zanstra temperature (see their Fig. 18). 
In this context, they showed that for optically thick objects, 
there is a good correlation between the two temperature determinations,
while for optically thin objects, the uncertainty exceeds 50\%.

Regarding luminosity, the uncertainty depends on the uncertainty in measuring the 5007\AA\  
flux and on the uncertainty in determining the distance to the host galaxy. 
However, the main source of uncertainty arises from the proportionality constant in 
Eq.~\ref{0.11}. While we can be certain that the uncertainty in this constant is large, 
assigning it a precise value remains purely speculative \citep{1992ApJ...389...27D}. 
For example, if we assume a large uncertainty for the proportionality constant (i.e. 50\%), 
then by propagating the errors we obtain
$\sigma(L/L_{\odot}) \approx 0.7 (L/L_{\odot})$.


\section{Discussion} \label{discusi}

As can be seen from Figs.~\ref{mw--lmc}, \ref{mw-5128} and \ref{mw-spy}, 
the MW sample of CSPNe is slightly cooler compared to the other galaxies. 
This difference can be explained by the fact that extragalactic 
PNe are generally detected through narrowband filters, particularly those centred on [O{\sc iii}]$\lambda$5007. 
Since the ionisation potential for this ion is 35.1~eV, using Eq.~\ref{txi}, 
we find that only extragalactic CSPNe with temperatures above 
$\approx$35,100~K can be detected, corresponding to $\log(T_{eff})>4.5$, which aligns with our findings.

As mentioned in Sec.~\ref{lmc}, except for the points with temperatures below 35100~K 
and the region corresponding to white dwarfs, the physical parameters of CSPNe in the LMC and SMC 
are distributed within the same region of the HR diagram as those of the MW. 
This is not the case for NGC~300 and NGC~5128 (Fig.~\ref{mw-5128}), 
whose objects are concentrated in a more restricted area.
In particular, in NGC~300, objects with log(L/L$_{\odot}$)<2.5 are likely underestimated, 
as this region corresponds to white dwarfs \citep[see Fig.~9 of][]{2020A&A...640A..10W}. 
In the case of NGC~5128, the clustering of points may be the result of systematic errors 
in the luminosity estimates, which are likely introduced by the assumptions made during their determination.

Concerning the effective temperature, according to Eq.~\ref{eq3}, 
when the EC exceeds a value of ten, we are likely dealing with objects hotter than $250 \times 10^3$~K. 
This temperature lies at the upper limit of the evolutionary models presented by 
\citet{2016A&A...588A..25M} for stellar masses greater than 0.8~M$_\odot$. 
Moreover, one of the hottest known stars, WR102, has an effective temperature of 210~kK 
\citep[see Table~4 of][]{2015A&A...581A.110T}. In this context, temperatures derived from 
EC values above ten should be treated with caution.


\section{Conclusions}
\label{conclu}

For the first time, the available data on extragalactic CSPNe have been analysed collectively. 
In addition to compiling information from the literature, physical parameters were computed for over 1000 objects. 
Perhaps the most paradigmatic case is that of the Magellanic Clouds, where, for the first time, 
the physical parameters of 168 known CSPNe --~gathered from 11 different papers~-- 
are presented together in an HR diagram.
This catalogue constitutes a necessary complement to the catalogue of Galactic CSPNe
\citep{2020A&A...640A..10W}.

In addition to the effective temperature and luminosity, we have included a 
handful of CSPNe with determined spectral types that have been reported in the literature. 
In fact, only 4\% of the total sample of 1073 objects have a known spectral type. 
Moreover, in some cases, only specific spectral features are reported, such as a P-Cygni 
profile or broad H$\alpha$ emission. The latter are part of the aforementioned 4\%.

Although the spectral types of only a few CSPNe have been accurately determined, 
it is clear that in the LMC, there are 11 objects with a [WC] central star. 
Of these, five are classified as [WC~5-8], indicating a relatively high proportion of intermediate 
subtypes compared to the distribution observed in our Galaxy \citep{2020A&A...640A..10W}.
While [WC] stars are easier to identify due to their strong emission lines \citep{2024A&A...686A..29W}, 
previous studies \citep[e.g.][that studied the PNe population of the Sagittarius 
dwarf spheroidal galaxy]{2006MNRAS.369..875Z} 
suggest that these spectral types are more frequent in metal-poor environments. 
Metallicities could also explain the presence of CSPNe of an intermediate subtype 
rarely found in the MW in the LMC \citep{1988MNRAS.234..583M}. 
In this sense, the SMC, which is more metal poor than the LMC (Table~\ref{gal-table}), 
should host a higher proportion of these spectral subtypes. However, 
only two [WC 5–8] detections have been reported in the SMC out of a total of 47 CSPNe, 
so no firm conclusions can be drawn given the small number of this sample. 
However, it is important to reiterate that the compiled catalogue is biased towards brighter stars, 
and therefore, drawing conclusions about relative population fractions based on metallicity remains highly uncertain.

Objects with high surface temperatures were identified in Magellanic Clouds and NGC~5128. 
If confirmed to be CSPNe, these sources would represent highly unusual cases due to their 
potentially high masses, and therefore warrant detailed follow-up with large telescopes. 
Such targeted observations could yield valuable data, offering new insights into stellar evolution.

In summary, the catalogue presented here is expected to serve as a valuable resource for guiding future studies, 
providing a comprehensive and homogeneous compilation of physical parameters for over a thousand CSPNe. 
It lays the groundwork for targeted spectroscopic follow-up, refined stellar evolution modelling, 
and comparative analyses across different galactic environments.


\begin{acknowledgements}

We are grateful to the anonymous referee for their valuable feedback, which greatly enhanced this manuscript.
    WW  thanks     Luis Gutiérrez-Soto  for his support.
     This work was partially supported by grants PICT 2021-00442 awarded by Foncyt, 
     grant PIP 2022 11220210100520CO by CONICET, 
     and by grant SeCyT UNC Consolidar project 2023  NRO: 33620230100103CB.
This research has made use of NASA’s Astrophysics Data System Service. 
This research has made use of the SIMBAD database, operated at CDS, Strasbourg, France. 
\end{acknowledgements}

The catalogue is only available in electronic form at the CDS via anonymous ftp to cdsarc.u-strasbg.fr (130.79.128.5) or via 
http://cdsweb.u-strasbg.fr/cgi-bin/qcat?J/A+A/.


%
%
\bibliographystyle{aa.bst} 
\bibliography{ref.bib} 



\begin{appendix}
\onecolumn
\begin{landscape}

\section{Catalogue of extragalactic CSPNe}



{\small

\begin{longtable}{l c l c r c r c c  c c c c c c}
\caption[Catalogue of extragalactic CSPNe]{Catalogue of extragalactic CSPNe. 
Description of the columns is in Section \ref{colum},
t.w. means this work.}
\label{cata} \\

\hline  \multicolumn{1}{c}{Name} & \multicolumn{1}{c}{galaxy} & \multicolumn{1}{c}{RA$_{J2000}$} & \multicolumn{1}{c}{DEC$_{J2000}$} & 
\multicolumn{1}{c}{log T$_{eff}$} & \multicolumn{1}{c}{Ref.} & \multicolumn{1}{c}{log(L/L{$_\odot$}) } & 
\multicolumn{1}{c}{Ref.} & \multicolumn{1}{c}{SpT} & 
\multicolumn{1}{c}{Ref.} & 
\multicolumn{1}{c}{RB}        &  \multicolumn{1}{c}{comm.}          \\ \hline 
\endfirsthead

\multicolumn{10}{c}%
{{\bfseries \tablename\ \thetable{} -- continued. }} \\
\hline  \multicolumn{1}{c}{Name} & \multicolumn{1}{c}{galaxy} & \multicolumn{1}{c}{RA$_{J2000}$} & \multicolumn{1}{c}{DEC$_{J2000}$} & 
\multicolumn{1}{c}{log T$_{eff}$} & \multicolumn{1}{c}{Ref.} & \multicolumn{1}{c}{log(L/L{$_\odot$})  } & 
\multicolumn{1}{c}{Ref.} & \multicolumn{1}{c}{SpT} & 
\multicolumn{1}{c}{Ref. } & \multicolumn{1}{c}{RB}          &  \multicolumn{1}{c}{comm.} \\ \hline 
\endhead

\hline\multicolumn{10}{r}{{Continued on next page}} \\ 
\endfoot

\hline \hline
\endlastfoot

DDHW 1 & Fornax & 02 39 49.30 & $-$34 32 47.00 &  4.74 & KG07 &  3.85 & KG07 & wels? & KG07 &  &  \\
x-x & IC 1613 &  &  &   &  &   &  & K  giant + ? & TR24 & TR24 &  \\
LMC-SMP 104A & LMC & 04 25 32.15 & $-$66 47 18.50 &  4.89 & VS07 &  4.69 & VS07 &  &  &  &  \\
LMC-SMP 1 & LMC & 04 38 34.79 & $-$70 36 43.28 &   &  &   &  & [WC 4] & MM23 &  &  \\
LMC-SMP 2 & LMC & 04 40 56.69 & $-$67 48 02.17 &  4.58 & HB04 &  3.42 & HB04 &  &  &  &  \\
LMC-SMP 3 & LMC & 04 42 24.00 & $-$66 13 01.66 &  4.67 & DM91c &  3.00 & DM91c &  &  &  &  \\
LMC-SMP 4 & LMC & 04 43 21.89 & $-$71 30 08.89 &  5.02 & TD24 &  3.81 & TD24 &  &  &  &  \\
LMC-SMP 6 & LMC & 04 47 38.98 & $-$72 28 20.57 &  5.15 & DM91c &  3.79 & DM91c &  &  &  &  \\
\end{longtable}
}
\end{landscape}





\section{Acronyms and bibliographical references }   
\label{acro}

BK13   \cite{2013ApJ...774....3B}

BV97   \cite{1997ApJ...480..290B}

DF85   \cite{1985ApJ...297..593D}

DM91   \cite{1991ApJ...367..115D}

DM91b  \cite{1991ApJ...374L..21D}

DM91c  \cite{1991ApJ...377..480D}

DV97   \cite{1997ApJ...474..188D}

FG18  \cite{2018ApJ...853...50F}

GH24   \cite{2024A&A...682A..70G}

GM07  \cite{2007MNRAS.375..715G}

GM22   \cite{2022A&A...657A..71G}

GM14   \cite{2014MNRAS.444.1705G}

H96    \cite{1996ASPC...96..127H}

HA00   \cite{2000JKAS...33...97H}

HB04   \cite{2004ApJ...611..294H}

HG14   \cite{2014A&A...561A...8H}

HP03   \cite{2003A&A...409..969H}

IM07   \cite{2007A&A...472..101I}

IM18   \cite{2018MNRAS.476.2605I}

JA23   \cite{2023MNRAS.523.2519J}

KG07   \cite{2007A&A...468..121K}

KL12   \cite{2012ApJ...753...12K}   

LR06   \cite{2006A&A...459..103L}

MB88   \cite{1988MNRAS.234..583M}

MH12   \cite{2012OAP....25..178M}

ML05   \cite{2005A&A...443..115M}

MM20   \cite{2020ApJ...898...85M}

MMb20   \cite{2020ApJ...888...54M}

MM23   \cite{2023ApJ...942...85M}

MN11   \cite{2011A&A...531A.157M}

MP04   \cite{2004A&A...426..779M}

OK16   \cite{2016MNRAS.462...12O}

PM04    \cite{2004RMxAC..20...41P}

PR97   \cite{1997A&A...324..674P}

PS07   \cite{2007A&A...476..745P}

SE09   \cite{2009A&A...494..515S}

SH03   \cite{2003ApJ...596..997S}

SL09   \cite{2009ApJ...702..733S}

SR23   \cite{Soemitro_2023A&A...671A.142S}

SS05   \cite{2005ApJ...622..294S}

TD24   \cite{2024A&A...688A..36T}

TR24   \cite{2024RNAAS...8...45T}

VS03   \cite{2003ApJ...597..298V}

VS04   \cite{2004ApJ...614..716V}

VS07   \cite{2007ApJ...656..831V}

W91    \cite{1991MNRAS.252P..47W}

ZG06   \cite{2006MNRAS.369..875Z}

ZV94   \cite{1994A&A...290..228Z}

\end{appendix}
\end{document}